\documentclass[twocolumn,prb,showpacs]{revtex4}
\usepackage{epsf,latexsym}
\newcommand{\up}{{\uparrow}}
\newcommand{\down}{{\downarrow}}

\newcommand{\be}{\begin{equation}}
\newcommand{\ee}{\end{equation}}

\newcommand{\bea}{\begin{eqnarray}}
\newcommand{\eea}{\end{eqnarray}}

\newcommand{\putfig}[2]{$$\leavevmode\hbox{\epsfxsize=#2 cm
   \epsffile{#1.eps}}$$}

\begin{document}
\topmargin-1.0cm
\ifx\undefined\psfig\def\psfig#1{    }\else\fi

\title  {Spin injection and electric
field effect in degenerate semiconductors}
\author  {Irene D'Amico}
\affiliation{ Istituto Nazionale per la Fisica della Materia (INFM);\\
Institute for Scientific Interchange, via Settimio Severo 65, 
I-10133 Torino,
Italy}
\date{\today}  
\begin{abstract}
We analyze spin-transport in semiconductors in the
regime characterized by $~~T\stackrel{<}{\sim}T_F$
(intermediate  to degenerate), where $T_F$ 
is the Fermi temperature.  Such a 
regime is of great importance since it includes
the lightly doped semiconductor structures used in most  experiments;  
we demonstrate that, at the same time,  it  corresponds to the regime in which
carrier-carrier interactions assume a relevant role.
Starting from a general formulation of the drift-diffusion equations,
which  includes many-body correlation effects, we perform detailed calculations
of the spin injection characteristics of various heterostructures,   
 and  analyze the combined effects of carrier density variation, applied
electric field  and Coulomb interaction.  We show the existence of  a {\it
degenerate} regime, peculiar to  semiconductors, which strongly differs, as
spin-transport is concerned, from the degenerate regime of metals.
\end{abstract}
\maketitle
\section{Introduction}
Despite many efforts and some substantial progress, major difficulties 
persist in the practical implementation of semiconductor
spintronics devices. 
One of the fundamental problems inherent the electrical 
injection of  spin currents into semiconductors is the so-called conductivity
mismatch\cite{schmidt:2000}: the large conductivity difference
between
metals and semiconductors induces a spin accumulation at the interface which in
turn drastically reduces spin injection into the semiconductor.

Recently Yu and Flatt\'e (YF)\cite{yu:2002}
 have pointed out the possibility of improving
the efficiency of spin injection through the application of a strong, but
experimentally accessible, electric
field. At variance with metals, in semiconductors,
 this field remains partially unscreened so that it is necessary
to  explicitly introduce  a  drift-related term in  the spin-transport
equations.

The YF analysis of the semiconductor high-field regime 
neglects electron-electron interaction effects -- aside from those responsible
for
macroscopic charge neutrality -- and, perhaps more importantly, focuses on the
high-field {\it non-degenerate} regime (see Eq.~(\ref{ndeg})),
especially insofar as the analysis of heterostructures are concerned. 

In the present paper we extend and refine their analysis in several directions. Starting from a more general formulation of the drift-diffusion equations,
which  includes many-body correlation effects, we perform detailed calculations
of the spin injection characteristics of various heterostructures.   

First of all we perform a careful analysis of the regime to which 
GaAs and  ZnSe-based structures (most commonly used in spintronics 
experiments) belong. We point out that, with the exception of lightly doped
samples at {\it room  temperature}, these systems are generally in a regime
which cannot be described by non-degenerate electron statistics.  In 
particular, the low temperature heterostructures used in
Refs.\onlinecite{Awshanature} and \onlinecite{schmidt:2002} are in the
degenerate regime (see Fig.~\ref{fig1}).
We then study  these structures under the effect of an applied electric field
and 
introduce the ``semiconductor degenerate regime''  characterized by  the
inequalities $k_BT\ll \varepsilon_F\ll e|E|L_s$, where $\varepsilon_F$ is the Fermi energy of the
carriers in the semiconductor and $E$ is the applied electric field.
  This high-field regime occurs, for comparable materials and
doping densities, at fields which
 are one order of magnitude smaller than in the
non-degenerate high-field regime. 
In particular, the use of non-degenerate statistics in heterostructures based
on 
GaAs, ZnSe or comparable materials, at a few Kelvin, can overestimate the effect of the
electric field by orders of magnitude (see Sec.IV).

As for Coulomb interactions, we demonstrate that the experimentally relevant
regime corresponds to the one in which the many body effects are at their
strongest. Neglecting these effects can induce both quantitative as well as
qualitative errors. Coulomb  interactions affect  the system
through the spin Coulomb drag effect\cite{SCD:2000, EPL:2001, PRB:2002}
and by reducing the longitudinal spin stiffness. In particular we will discuss
how the spin drag explicitly modifies spin-transport whenever 
materials with different magnetic properties are joined.

 The numerical part of this paper concentrates on sandwiched structures
(ferromagnet (FM)/non-magnetic semiconductor (NMS)/ferromagnet) which are of
 the utmost
importance in device applications based on the magnetoresistance effect.  We
have performed a detailed analysis  of the spatial dependence of various
quantities such as electrochemical potentials, density and current 
polarizations for both parallel and antiparallel
alignment of the ferromagnets.   
At variance with the parallel case, we find that in the antiparallel case, the
magnetoresistance can be
 {\it increased} by a strong electric field.   

Another important issue we  address is the effect of the
dependence of the  spin diffusion constant $D_s$ on
carrier density when $k_BT\stackrel{<}{\sim}\varepsilon_F$ (see inset of
Fig.~\ref{fig1}). We will show that, 
due to this dependence, it is possible to greatly
enhance the local spin current density 
by carefully choosing the system parameters.

This paper is organized as follows: in Sec. II we discuss the system regimes as
doping density, temperature, and electric field are varied;
in Sec. III we derive the drift-diffusion equation for spin-transport including
both the Coulomb interaction and the electric field, and discuss the behavior
of the upstream and downstream diffusion lengths; in Sec. IV we discuss the limits of the non-degenerate approximation;
in Sec. V we consider the
spin current and the implications of varying the carrier
density on this quantity; in Sections  VI  and VII we analyze
 FM/NMS  and FM/NMS/FM  heterostructures, respectively;   in Sec. VIII we
describe the effects of Coulomb interactions;  finally
in Sec. IX  we draw our main  conclusions. Two appendices, discussing 
some details of the proposed heterostructure-related formalism
  and proposing ``exact''
  equations for describing  spin transport in
non-interacting degenerate heterostructures  conclude the paper.

\section{Doping density, temperature and electric field -related regimes} 
This work focuses on electrically driven spin transport: 
due to the presence of the external 
electric field, the system is characterized by {\it three different energy scales}
whose relative magnitude determines the system behavior.
Such energies are the Fermi energy $\varepsilon_F=\hbar^2 k_F^2/2m$ (which strongly depends on the carrier density), the thermal energy $k_BT$ 
and the energy related to the electric field $e|E|L_s$, where 
$L_s=\sqrt{D_s\tau_s}$ is the diffusion length, $D_s$ the diffusion constant and $\tau_s$ the spin-flip time.

\begin{figure}
\putfig{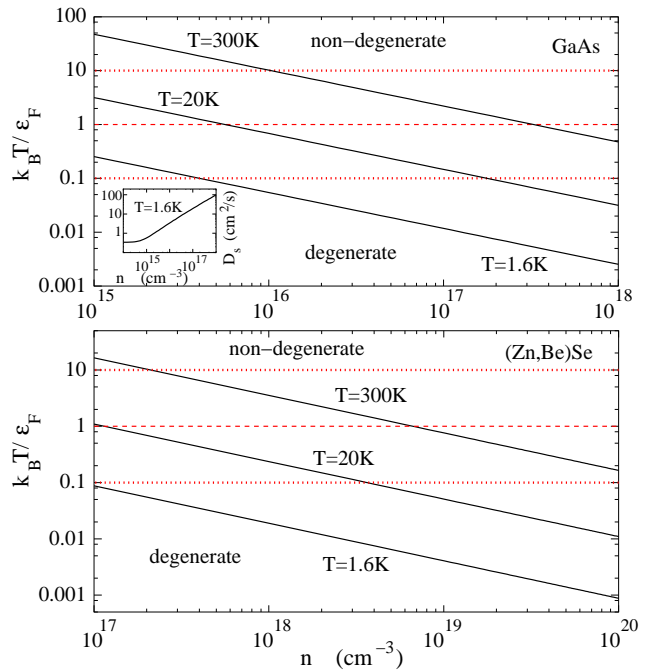}{8.5}
\caption{ $k_BT/\varepsilon_F$ vs carrier density $n$ for GaAs (upper panel, $m=0.067m_0$, $\epsilon=12$)  and (Zn, Be)Se (lower panel, $m=0.5m_0$, $\epsilon=12$) parameters, at  $T=1.6$K, $20$K, $300$K, as 
labelled. Degenerate and non-degenerate regions are indicated as well.
Inset: $D_s$  vs carrier density for $T=1.6$K and GaAs parameters.
}
\label{fig1}
\end{figure}

Let us first of all define the regimes the system 
undergoes depending on the relation between thermal and Fermi energies
\bea 
k_BT &\gg& \varepsilon_F ~\mbox{non-degenerate regime}\label{ndeg}\\ 
k_BT &\ll& \varepsilon_F ~\mbox{degenerate regime}\label{deg}\\
k_BT &\approx & \varepsilon_F ~\mbox{intermediate regime}.\label{inter}
\eea
The intermediate regime corresponds to the crossover region between the first two, roughly the region between the dotted lines in  Fig.~\ref{fig1}.

We will concentrate our discussion on NMS in the {\it intermediate and degenerate} regimes. Such regimes are of utmost importance since they 
include most of the semiconductor structures used in spintronics related 
experiments, such as n-doped GaAs\cite{Awshanature} or 
Zn$_{0.97}$Be$_{0.03}$Se\cite{schmidt:2002} ones.
This is illustrated in
 Fig.~\ref{fig1}, where we plot the ratio  $k_BT/\varepsilon_F$ as a function of the carrier density $n$ for  n-doped GaAs (upper panel) and (Zn,Be)Se (lower
panel).
Three different values of the temperature are considered (as labelled). The regions corresponding to degenerate and non-degenerate systems are
specified as well. 
As can be seen, up to room temperatures, for densities as low as
$n\sim 5\cdot 10^{16}$cm$^{-3}$ in GaAs or $n\sim 10^{18}$cm$^{-3}$ in (Zn,Be)Se the system is at most in the intermediate regime; 
for $T=1.6$K, 
a temperature often used in experiments, the system remains in the {\it degenerate} regime   for the whole relevant density range.

 Since, as just discussed,  in most experimental situations $k_BT\stackrel{<}{\sim}\varepsilon_F$, the other relevant information to determine the expected system behavior is contained  in the ratio $e|E|L_s/\varepsilon_F$.

\begin{figure}
\putfig{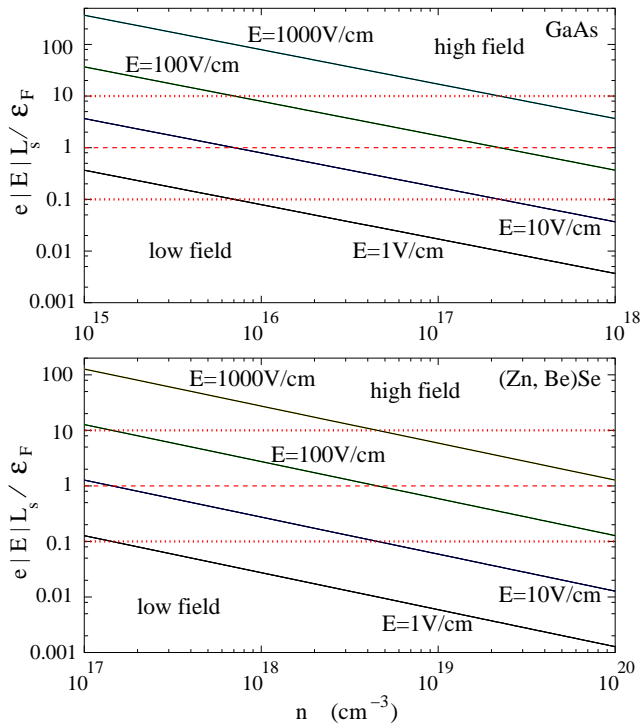}{8.5}
\caption{$e|E|Ls/\varepsilon_F$ vs carrier density $n$ for GaAs (upper panel)  and (Zn, Be)Se (lower panel) parameters, for  $E=1,~10,~100,~1000$V/cm as 
labelled. High and low field regions are indicated as well.
}
\label{fig2.1}
\end{figure}

In Fig.~\ref{fig2.1} we plot this ratio  as a function of density and for fields up to 1000 V/cm, for both GaAs and (Zn,Be)Se. In both cases we have chosen $L_s=2\mu$m. We will consider this same value in all the calculations
 presented\cite{note3.1}. 
The high ($e|E|L_s\gg\varepsilon_F$) and low ($e|E|L_s\ll\varepsilon_F$) field regions are explicitly indicated.
We notice that, even at very low densities, it is necessary to apply fields
of the order of 100 V/cm to enter the high field regime.

As we will explain in detail, the comparison of  the energy scales, as done in Fig.~\ref{fig1} and  Fig.~\ref{fig2.1}, 
 is a very efficient way to understand in which regime the considered system is and deduce how and if the electric field is going to affect the diffusion lengths and the spin-transport related quantities. 
In particular
  {\it at variance with metals}, in semiconductors  the Fermi energy can be so low that, even for moderate electric fields, we can fulfill the condition  $e|E|L_s\gg\varepsilon_F\gg k_BT$. This is what we define as ``semiconductor degenerate regime'' to distinguish it from the well-know ``metallic'' degenerate regime in which   $\varepsilon_F\gg e|E|L_s$ for any reasonable field. 
In the ``semiconductor degenerate regime'' 
 the electric field heavily affects spin-transport even if the system is
 degenerate.
In order to illustrate this point, we will analyze in detail
the behavior of the penetration lengths $L_d$ and $L_u$ as electric field 
and carrier density change.

Due to its importance, in this article we will  mainly 
discuss
 n-doped GaAs at $T=1.6$K; our results are based  on
 the interplay among different energy scales, so that they can be readily 
extended to  other 
semiconductor materials.

\section{Interacting drift-diffusion equation}
First of all we will sketch the derivation of the general
 drift-diffusion equation for spin-transport which {\it includes Coulomb 
interactions} among carriers. We 
 refer the reader to Ref.~\onlinecite{PRB:2002} for the details. 

We  start from the
drift diffusion equation for the interacting spin-$\alpha$ charge current 
\begin{equation} \label{eq1} \vec j_\alpha (\vec r)=
\sum_\beta \left(\sigma_{\alpha \beta} \vec E +eD_{\alpha \beta}\vec \nabla 
n_\beta\right),
\end{equation}
where    $\sigma_{\alpha \beta}$ is the ({\it non-diagonal}) homogeneous conductivity matrix of the electron gas and
 $D_{\alpha \beta}$ is
the ({\it non-diagonal}) diffusion matrix. 
Throughout the paper Greek indices will correspond to  the spin-variables $\uparrow,\downarrow$.
We substitute Eq.~(\ref{eq1}) in the generalized continuity equation
for
the spin-density components 
\be
{\partial \Delta n_\alpha(\vec r,t) \over \partial t} = - {\Delta n_\alpha
(\vec r,t) \over \tau_{sf, \alpha}} + {\Delta n_{\bar{\alpha}}
(\vec r,t) \over \tau_{ sf, \bar \alpha}} +{\vec \nabla \cdot \vec
j_\alpha(\vec r)\over e},
\ee
 where 
$\Delta n_\alpha( \vec r ,t) \equiv
n_\alpha(\vec r,t) -n_\alpha^{(0)}$, $n_\alpha(\vec r,t)$ is the $\alpha$
density component, $n_\alpha^{(0)}$ its equilibrium
 value  and  
$\tau_{sf, \alpha}$ is the $\alpha$
spin-flip relaxation time. Carefully applying\cite{PRB:2002} 
the local charge neutrality constraint 
\begin{equation}\Delta n_{\uparrow}
(r) = - \Delta n_{\downarrow} (r)\label{chneutr}
\end{equation}
and considering the regime linear in $\nabla n_\alpha$,
the general interacting drift-diffusion equation is 
 obtained\cite{PRB:2002}.

Finally, imposing the steady state condition ${\partial (\Delta n_{\uparrow}-\Delta n_{\downarrow}) / \partial t} =0$ we obtain the steady-state {\it interacting} drift-diffusion equation
 \bea \label{steady}  & &- {(\Delta n_{\uparrow}-\Delta n_{\downarrow})  \over \tau_s} + D_s
\nabla^2 (\Delta n_{\uparrow}-\Delta n_{\downarrow})\nonumber \\ & & + \mu_s \vec E \cdot \vec \nabla (\Delta n_{\uparrow}-\Delta n_{\downarrow})=0,
 \eea 
where  $\tau_s=(1/\tau_{sf, \uparrow
} +1/\tau_{sf, \downarrow})^{-1}$ is the spin relaxation
time, and the effective interacting
mobility and diffusion constants are given, for a non-magnetic system\cite{note}, by
\begin{eqnarray} \label{muspin}
 \mu_s &=& e
\tau_D / m
\end{eqnarray} 
and
\begin{eqnarray} \label{Dspin}
 D_s &=& {\mu_s  \over e} S n {1 \over 1 - \rho_{\uparrow
\downarrow}/\rho_D}.
\end{eqnarray}  
Here $\rho_D =m/ ne^2 \tau_D$ is the ordinary Drude resistivity, $m$ the effective mass of the carriers,
$S$ the (interacting) static longitudinal spin-stiffness\cite{PRB:2002}, $n= n_{\uparrow}+n_{\downarrow}$ the carrier density
 and $\rho_{\uparrow
\downarrow}$ the spin-transresistivity, which measures the momentum rate exchanged between spin up and spin down carriers\cite{SCD:2000,EPL:2001,PRB:2002}.
Eq.~(\ref{steady}) is valid independently of the value
of $k_BT/\varepsilon_F$.

If the sample is homogeneous and $\vec E\parallel\hat{x}$
the solution to Eq.~(\ref{steady}) is given by
 \begin{equation}\Delta n_{\uparrow}-\Delta n_{\downarrow}=Ae^{x/L_u}+Be^{-x/L_d}\label{deltam}
\end{equation} with  A, B  determined by the boundary conditions
and
 \begin{equation}\label{Lud} L_{u,d}^{-1}=\pm {\mu_s|E|\over 2D_s}+
\sqrt{ \left({\mu_s|E|\over 2D_s}\right)^2+{1\over D_s\tau_s}}.
\end{equation} 
$L_{u,d}$ are  the {\it interacting} upstream and downstream
``penetration'' lengths\cite{note3}. 
Such quantities correspond to the average decay lengths of an injected spin-unbalance in the upstream ($L_u$) and downstream ($L_d$) directions. 

\begin{figure}
\putfig{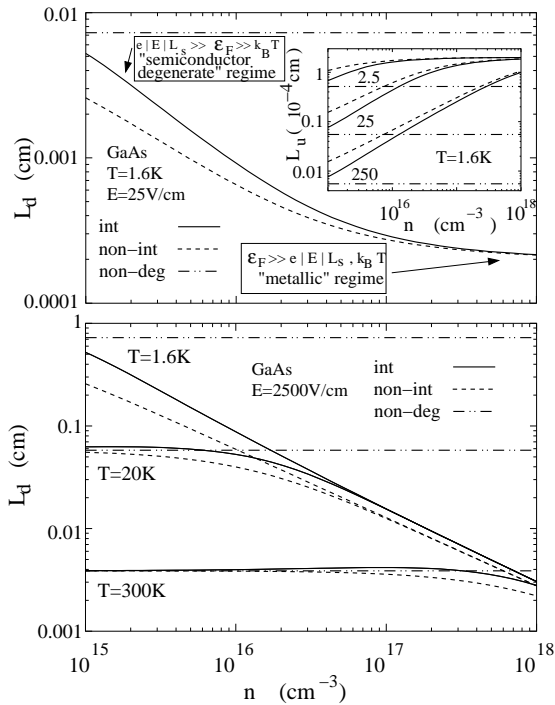}{7.5}
\caption{Upper panel: downstream diffusion length $L_d$ vs carrier density for GaAs parameters, T=1.6K and E=25V/cm. Solid line refers to the  interacting calculation, dashed line to the non-interacting approximation and dashed-double dotted line to the non-degenerate limit. Inset: Upstream diffusion length $L_u$ vs density $n$ for GaAs parameters at $T=1.6$K and three different fields (as labelled in V/cm units). Line types as in the main panel. 
Lower panel: downstream diffusion length $L_d$ vs carrier density for GaAs parameters, T=1.6K, 20K  and 300K (as labelled) and E=2500V/cm. Line types as in the upper panel.
}
\label{fig2}
\end{figure}

The upper panel of Fig.~\ref{fig2} shows the behavior of $L_{d}$,    
as a function of density for GaAs at T=1.6K and  $E=25$V/cm. The solid line 
represents the fully interacting calculation, the dashed one the 
non-interacting approximation and the dashed double dot line the 
non-degenerate limit obtained by setting $D_s=\mu_sk_BT/e$ in Eq.~(\ref{Lud}).
In this, and in all the following calculations, we have set $\mu_s=3\times 10^3$cm$^2$/Vs.\cite{Awshanature} However we stress that, at least in the degenerate regime, for which usually $\rho_{\uparrow
\downarrow}/\rho_D\ll 1$, $L_{u,d}$ is basically independent from $\mu_s$ (see Eqs.~(\ref{Lud}) and (\ref{Dspin})).

Let us focus on  the crossovers undergone by
 the system  when the carrier density is increased. 
At this temperature even at very low densities the system is degenerate ($\varepsilon_F>>k_BT$), so the interesting interplay is 
 between the Fermi energy and the electric field related energy: 
as shown in the figure, depending on the system density,
  the Fermi energy can be low enough to enter, even for moderate electric fields, the ``semiconductor degenerate regime'' $e|E|L_s>>\varepsilon_F$. 
 The existence of such a regime implies that, {\it even for a degenerate system}, the drift term in Eq.~(\ref{steady}) cannot be neglected, i.e. it is not possible to describe spin-transport using a diffusion equation for the electrochemical potentials, as usually done for metals\cite{vanson:1987}.
Even in {\it degenerate} semiconductor systems in fact, the drift term 
severely modifies the  penetration lengths $L_{u,d}$, varying their values over order of magnitudes (as clearly shown for $L_d$
in Fig.~\ref{fig2}).
As a rule of thumb, we see that in this regime the order of magnitude of such a variation is set by the ratio $e|E|L_s/\varepsilon_F$.
The ``semiconductor degenerate'' regime extends to higher densities when the applied field is increased.
Notice that, even in this regime, the actual value of $L_d$ strongly
differs from  its non-degenerate limit. 
By  increasing the carrier density, 
 the system will eventually enter the ``metallic'' regime (as indicated in 
 the figure), in which the drift term can  be neglected and $L_{u,d}$ recover their unperturbed value $L_s$.

Similar results, i.e. order of magnitude variations depending on the carrier density and
quantitative importance of Coulomb interactions   are 
obtained for the upstream penetration length $L_u$ (see upper panel inset),
 with the key difference that in the ``semiconductor degenerate regime''  
$L_u$ is strongly {\it reduced} by the electric field effect, being the 
product  $L_u\cdot L_d=L_s$ constant\cite{note3.1}.  

It is interesting to ask what happens in the high field regime $e|E|L_s>>(\varepsilon_F,~k_BT)$ when both temperature and carrier density varies. 
In  Fig.~\ref{fig2}, lower panel, we plot $L_d$ in the high field regime 
(E=2500 V/cm),  
as a function of carrier density and for three different temperatures, $T=1.6$K, $T=20$K and $T=300$K. In addition to the crossovers discussed in relation to the upper panel,
  for low densities and $T=300$K the system enters the non-degenerate regime $\varepsilon_F<<k_BT$ (see also Fig.~\ref{fig1}).
In this region in fact, the non-degenerate approximation coincides with the interacting calculation.
 Fig.~\ref{fig2} shows that the greatest variations due to the presence of the electric field are actually reached by lowering the temperature and 
for any reasonable density and the materials considered, correspond to  the 
``semiconductor degenerate'' regime.

\section{Limits of the non-degenerate approximation}
As demonstrated in Sec. II, the use of non-degenerate electron statistics 
to describe  carrier dynamics in commonly-used semiconductors is justified only
for very lightly doped semiconductors  at high temperatures. 
For this reason, when  analyzing spin transport, particular 
care must be taken in determining  the correct system regime:
for some other problems  an easy-to-implement 
approximation, used outside its supposed
validity regime, can still yield reasonable 
estimates\cite{noteDFT}; unfortunately this is not the case for 
the  non-degenerate approximation  in semiconductor spin-transport. 
It is tempting, 
since it leads to simpler analytical formulas, to approximate 
 $D_s$  in the various quantities (from the penetration lengths to the spin current, from the injected density
 polarization to the magnetoresistance), by its non-degenerate, classical expression.
Unfortunately as it has been experimentally shown\cite{Awshanature} and 
theoretically
 confirmed\cite{flatte:2000}, 
the spin diffusion constant varies over {\it orders of magnitude}
 going from the non-degenerate 
to the intermediate and degenerate regime (see  inset of Fig.~\ref{fig1}, in which $D_s$ is plotted as a function of $n$
 for GaAs parameters). 
 This implies that approximating spin transport behavior by   non-degenerate 
expressions in the wrong regime can lead to quite large errors.
In Fig.~\ref{fig2}, Fig.~\ref{fig5} and Fig.~\ref{fig14}  we have 
compared the interacting calculation (solid line) with its non-degenerate approximation (dashed double-dot line) for some of the most relevant quantities, as
  penetration lengths, spin current, and magnetoresistance respectively.
  At low temperatures, even for densities as low as $n\sim 5\times 10^{16}$cm$^{-3}$ and fields as small as E=10V/cm (low-field regime), 
this approximation can overestimate
 the field effect by {\it orders of magnitude}.
 Such a discrepancy is quite general and is
 found also for quantities such as injected density and current polarizations
(not shown). 

\section{The spin current}
We will now focus on the spin-current in a non-magnetic system.
Starting from the drift-diffusion Eq.~(\ref{eq1}), and imposing the conditions
$D_s=D_{\uparrow\uparrow}-D_{\uparrow\downarrow}=D_{\downarrow\downarrow}-
D_{\downarrow\uparrow}$ and $\mu_s=\mu_{\uparrow\uparrow}+
\mu_{\uparrow\downarrow}=\mu_{\downarrow\downarrow}+\mu_{\downarrow\uparrow}$ which are obtained from the general  formula for $D_s$ and 
$\mu_s$\cite{PRB:2002} in the non-magnetic limit and in the regime  linear in 
$\nabla n_\alpha$, 
the charge current becomes
\begin{equation} \label{ccu}
 \vec j (\vec r):= j_\uparrow (\vec r)+j_\downarrow (\vec r)=
e\vec E \mu_s( n_\uparrow+ n_\downarrow)
\end{equation} while
the spin current assumes the form:
\be \label{scu1}
 \vec j_s (\vec r) :=  j_\uparrow (\vec r)- j_\downarrow (\vec r)= \vec jP(x)+eD_sn\nabla P(x)
\ee with $P(x)=( n_\uparrow- n_\downarrow)/n$ the density
polarization.
If we  consider a spin unbalance  injected at $x=0$, i.e. such that in Eq.~(\ref{deltam}) either $A=0$ ($x>0$) or $B=0$ ($x<0$), $\Delta n_{\uparrow}-\Delta n_{\downarrow}$
 will acquire a simple exponential 
form; the upstream ($x<0$) (downstream ($x>0$)) spin current will then be given by
\be 
 \vec j_s^{u,d} (\vec r)=
\left[\vec j \mp (-e){nD_s\over L_{u,d}}\hat{x}\right]P(x)
=(\vec j +\vec j_D^{u,d})P(x).\label{scu}
\ee
Eq.~(\ref{scu}) is very interesting because it clearly shows how the spin current is composed: $\vec j_s^{u,d}$ results in fact  as the sum of two distinct parts, the first corresponding to the {\it total} drift current $\vec j$ and the second being a {\it total}  diffusion current $\vec j_D$. Both the components are 
``weighted'' by $P(x)$, 
the percentage of spin-polarized carriers. It is interesting to notice that in the downstream case, $\vec j\parallel \vec j_D$, so that both diffusion and drift phenomena positively contribute to the spin-current, while in the upstream case $\vec j$ and $\vec j_D$ are in competition. It is easy to see that in the limit of very large electric fields, $1/L_d\to 0$ so that 
$\vec j_s^{d} (\vec r)=\vec jP(x)$, while $1/L_u\to |E|\mu_s/D_s$ and $\vec j_s^{u} (\vec r)$ exactly vanishes. 
 Keeping in mind that the diffusion constant increases by orders of magnitude when going from the non-degenerate to the degenerate regime, 
Eq.~(\ref{scu}) suggests that a method to increment the spin current for a given electric field should be to enter the degenerate regime (for example by lowering the temperature), increasing in this way the diffusive part of the current. If instead  
the doping (or the temperature) is kept constant, an increase in the electric field will increase the drift current
$\vec j$. In this last case though, 
$1/L_d$ will decrease, and to  the incremented  $\vec j$ 
 will correspond  a  decreasing of $\vec j_D^d$.

\begin{figure}
\putfig{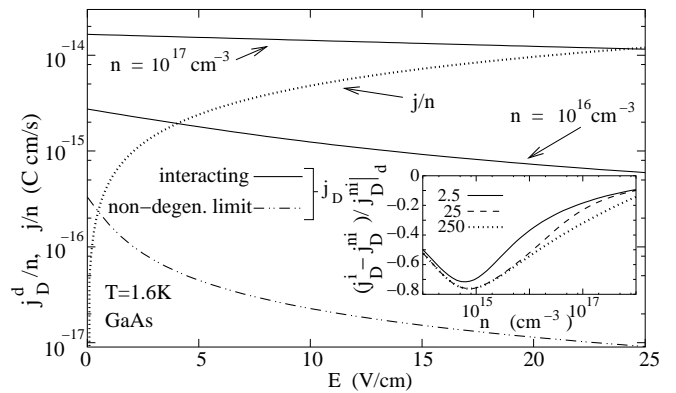}{8.8}
\caption{Diffusion current density $j_D$ and total current density $j$ (rescaled by total carrier density $n$) vs electric field $E$  in the downstream direction for GaAs parameters, T=1.6K and two different carrier densities (as labelled). Solid lines refer to the interacting $j_D/n$,
 and the dashed-double-dot line to the non-degenerate limit of $j_D/n$.
Dotted line refers to $j/n$.
Inset: correction to the non-interacting approximation $(j_D^i-j_D^{ni})/j_D^{ni}$ vs carrier density for $E=2.5,~25,~250$V/cm. 
}
\label{fig5}
\end{figure}

We illustrate this behavior in  Fig.~\ref{fig5}. In this figure we plot
$j_D^d$ and $j$ at $T=1.6$K rescaled by the trivial (and common) dependence on the carrier density $n$ as a function of the applied electric field 
 and for two different densities (as indicated).
$j/n$ (dotted curve) does not depend on the density nor on Coulomb interaction.
The fully interacting $j_D^d/n$ corresponds to the solid line.
If we fix the density and increase the electric field, $j_D$ 
decreases and $j$ increases; but if we fix the electric field (considering a 
moderate electric field), and increase the density, $j_D$ 
increases while $j$ does not change, so that there is no competition between the two quantities.
 We underline once more that this is not due to a trivial
linear scaling with the carrier density, but to the behavior of the diffusion constant when entering the degenerate regime.    
Regarding $P(x)$, the second factor in Eq.~(\ref{scu}),  if we  
consider as injector a not-fully polarized metal 
in the degenerate regime and for moderate fields, we can easily deduce
from Eq.~(\ref{P0}) (next session), that
 $P(0)\sim n^{1/3}$, i.e. 
$P(x)$ slightly increases with carrier density. 
This implies that in such a regime 
our conclusions hold for the full spin current $j_s^d$.

\section{FM/NMS junction}
\begin{figure}
\putfig{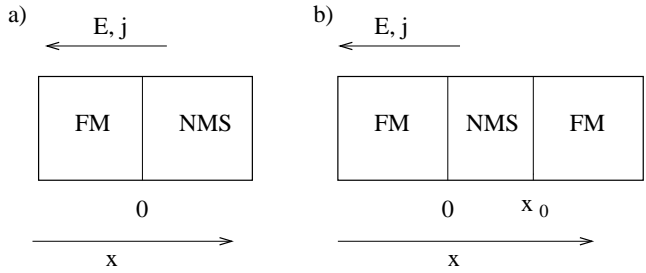}{8.5}
\caption{Schematic view of the FM/NMS (left) and FM/NMS/FM (right) heterostructures considered.
}
\label{fig6.1}
\end{figure}

We will now concentrate on  FM/NMS heterostructures. 
The schematic view of the system is illustrated in Fig.~\ref{fig6.1}a.
 We will first derive explicit expressions for the quantities of interest (electrochemical potential, 
density and current polarization, magnetoresistance, etc)
in the  interacting case, for {\it general} carrier density $n$ and temperature $T$  and then concentrate on 
 the  degenerate regime.
As underlined before, this regime is experimentally relevant, and, as discussed in detail in Sec. VII,  in it
 Coulomb interactions are quantitatively important.

In the {\it non-interacting} approximation it is possible to write explicit expressions for the chemical potential in both the non-degenerate and degenerate regime. By using them, the equations for the quantities of interest can be derived (see respectively Ref.~\onlinecite{yu:2002} and Appendix B). Since we want to derive instead equations which include Coulomb interactions and are valid independently from the doping regime,  
let us  consider the Taylor expansion of the interacting chemical potential
 \begin{equation}
\mu_\sigma^{chem}(n_\uparrow,n_\downarrow)=\mu_{0\sigma}^{chem}+
{\partial\mu_\sigma^{chem}\over\partial n_\sigma}\Delta n_\sigma
+{\partial\mu_\sigma^{chem}\over\partial n_{\bar{\sigma}}}\Delta n_{\bar{\sigma}}+.....\label{Taylor}
\end{equation}
If we use the relation
\begin{equation}S_{\sigma\delta}\equiv {\partial\mu_\sigma^{chem}\over\partial n_\delta}
\end{equation} between 
the $\{\sigma\delta\}$ element of the spin-stiffness tensor and the chemical potential,
and the local charge neutrality constraint Eq.~(\ref{chneutr}), we 
can rewrite such a potential as
\be
\mu_\sigma^{chem} \approx \mu_{0\sigma}^{chem} +
(S_{\sigma\sigma}-S_{\sigma\bar{\sigma}})\Delta n_\sigma.\label{muappr}
\ee
The limits of this approximation are discussed in Appendix A. 
The excess 
electrochemical potential  associated to Eq.~(\ref{muappr}) then becomes
\bea
\Delta \mu_\sigma &\equiv& \mu_\sigma-\mu_{0\sigma}^{chem} \label{deltamugen}
\\&\approx&(S_{\sigma\sigma}-S_{\sigma\bar{\sigma}})\Delta n_\sigma+e\vec{E}\cdot \vec{x}+B.\label{deltamu}
\eea
Eq.~(\ref{deltamu}) is general and is valid both for polarized and non-magnetic
 materials.

By using  the relation between the spin stiffness and the spin stiffness tensor components 
\begin{equation} S=(S_{\uparrow\uparrow}-S_{\uparrow\downarrow}+S_{\downarrow\downarrow}-S_{\downarrow\uparrow})/4,\end{equation}
the symmetries in the NMS  ($S_{\uparrow\uparrow}-S_{\uparrow\downarrow}=S_{\downarrow\downarrow}-S_{\downarrow\uparrow}$), and
the local charge neutrality constraint,
the electrochemical potential in the NMS ($x>0$) is given by
\bea
\Delta \mu_{\up(\down)}  &=& \pm SnP(x)
+e\vec{E}\cdot \vec{x}+B\nonumber \\
  &=& \pm {D_se\over\mu_s}(1-\rho_{\uparrow\downarrow}/\rho_D)P(x)
+e\vec{E}\cdot \vec{x}+B,\label{deltamuNMS}
\eea
where the plus (minus) sign refers to  the $\up$ ($\down$) spin component and
 we have used Eq.~(\ref{Dspin}) in the second expression.

If we assume that the FM  (diluted magnetic semiconductor or metal) is in the ``metallic'' degenerate regime (as it is often the case), the electrochemical potential on the FM side ($x<0$) will be\cite{yu:2002,vignale:2003}
\begin{equation}
\left(\begin{array}{c}\Delta \mu_\uparrow \\\Delta \mu_\downarrow
\end{array}\right)=e{\vec{j}\cdot\vec{x}\over\sigma^f} \left(\begin{array}{c}1\\1\end{array}\right)+Ce\vec{j}\cdot\hat{x}\left(\begin{array}{c}1/\sigma^f_\uparrow\\-1/\sigma^f_\downarrow\end{array}\right)\exp({x\over L_f})
\end{equation}
In this expression Coulomb interactions are included through $\sigma^f_\sigma\equiv\sigma^f_{\sigma\sigma}+\sigma^f_{\sigma\bar{\sigma}}$\cite{vignale:2003}. Since in this regime the spin transresistivity is much smaller than the Drude resistivity\cite{SCD:2000}, we will neglect the former
in our calculations.

If we assume transparent interfaces,
the boundary conditions to be satisfied at $x=0$ will be
\begin{eqnarray}
\left.j_s\right|_{0^-}&=&\left.j_s\right|_{0^+} \label{cond1}\\
\left.\Delta \mu_\uparrow\right|_{0^-}&=&\left.\Delta \mu_\uparrow\right|_{0^+}\label{cond2} \\
\left.\Delta \mu_\downarrow\right|_{0^-}&=&\left.\Delta \mu_\downarrow\right|_{0^+}.\label{cond3}
\end{eqnarray}
 The spin current components are given by
\begin{equation}e\vec{j}_\sigma=\sum_\delta \sigma_{\sigma\delta}{\partial\Delta\mu_\delta\over\partial x}\hat{x},
\end{equation}
and the spin currents in the two materials will then become
\begin{eqnarray}
\vec{j}_s^f&=&\vec{j}\left(p_f+{2C\over L_f}\exp({x\over L_f})\right)\\
\vec{j}_s^{sc}&=&P(x)(\vec{j}+\vec{j}_D^d)\label{jNMS}
\end{eqnarray} where the indexes $f,~sc$ correspond respectively to the FM and to the NMS,
$p_f\equiv(\sigma_\uparrow^f-\sigma_\downarrow^f)/(\sigma_\uparrow^f+\sigma_\downarrow^f)$ and $\vec{j}_D^d$ is defined by Eq.~(\ref{scu}).

By using Eqs.~(\ref{deltamuNMS})-(\ref{jNMS}), after some straightforward
algebra, 
 we derive the expressions for the quantities of interest. 

 The electrochemical potential in the NMS  will  be
\be 
\Delta \mu_{\up(\down)}  = -SnP(0)[p_f\mp\exp(-x/L_d)]  +e\vec{E}\cdot \vec{x},
\label{NMSep}\ee
where the - (+) sign refers to the $\up$ ($\down$) spin component.
In particular the expression for the interface band splitting is given by
\be 
\Delta \mu_{\up}(0)-\Delta\mu_{\down}(0)  = 2SnP(0)\stackrel{E\to\infty}{=}2Snp_f 
\ee
We see  that the splitting saturates for high electric fields.
The relation $Sn=D_se(1-\rho_{\uparrow\downarrow}/\rho_D)/\mu_s$ implies on the other hand that
 this limit
can be increased by increasing the diffusion constant $D_s$ (for example by
 increasing the system density
as previously discussed).

The magnetoresistance (which is related to the electrochemical potential sum at the interface and will be discussed in detail later) can be written as
 \be 
R_m=-{\Delta \mu_{\up}(0)+\Delta\mu_{\down}(0)\over2e|j|}  = {SnP(0)p_f\over e|j|}.\label{Rm1jct0}
\ee 
The equation for the injected density polarization is
\begin{equation}P(0)=L_u\left({1\over L_u}-{1\over L_d}\right){p_f\over
{\sigma_f\over\sigma_{sc}}{(1-p_f^2)\over L_f}L_u(1-\rho_{\uparrow\downarrow}/\rho_D)+1}\label{P0}
\end{equation}
while the one for the injected current polarization is
\begin{equation}\alpha(0)={p_f\over
{\sigma_f\over\sigma_{sc}}{(1-p_f^2)\over L_f}L_u(1-\rho_{\uparrow\downarrow}/\rho_D)+1},\label{alpha0}\end{equation}
where we have used the relations $E\mu_s/D_s=(1/L_u-1/L_d)$.
We stress that Eqs.~(\ref{NMSep})-(\ref{alpha0}) are valid in all density and field regimes and represent the fully interacting expressions. 
 
\begin{figure}
\putfig{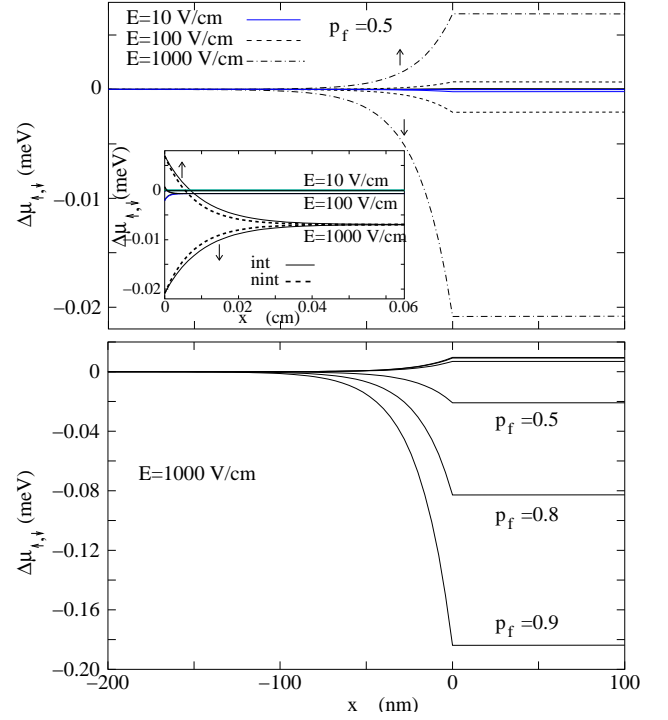}{8.5}
\caption{Upper panel:Up and down electrochemical potentials (without the term linear in $x$) at the junction vs $x$ for three different fields (as labelled) and $p_f=0.5$. Inset: as in main panel but on the NMS side of the junction. The non-interacting approximation (dashed line) is plotted as well.
Lower panel: As in the upper panel but
 for three different FM polarizations (as labelled) and $E=1000$V/cm.
}
\label{fig8}
\end{figure}

Fig.~\ref{fig8} 
 illustrates the behavior of the $\up$ and $\down$ 
electrochemical potentials (excluding the trivial linear dependence on $x$, as
 we will do in all  figures related to such quantities) as a function of $x$
across the FM/NMS junction  and for increasing field and FM 
polarization.
 The  parameters used in the calculations we present are $n=5.2\cdot 10^{16}$cm$^{-3}$ ($\varepsilon_F=7.6$meV), $\sigma^f/\sigma_{sc}=95$, $L_f=20$nm, $L_s=2\mu$m\cite{note3.1}.
In the upper panel of Fig.~\ref{fig8} the FM polarization is $p_f=0.5$.
 As the figure shows, the splitting of the FM sub-bands at the interface increases asymmetrically in respect to their asymptotic values (at $-\infty$), with the electric field due to the difference between the conductivities
$\sigma_\up^f,~\sigma_\down^f$. In the inset 
 we plot the same electrochemical potentials but on the NMS side of the junction ($x>0$). 
As expected the potentials are now symmetric with respect to their asymptotic
value  (at $\infty$) and the extension of the spin-polarized region in which $\nabla \mu_\sigma\ne const$
 extends in space as the electric field increases. As can be seen, 
the potential drop at the interface (the difference between the values of the electrochemical potentials at $\pm \infty$), which is related to the magnetoresistance, increases with the field.

\begin{figure}
\putfig{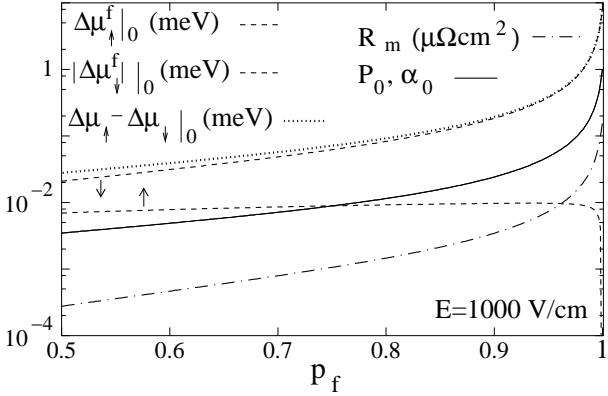}{8.0}
\caption{Up and down electrochemical potentials (dashed lines), their difference (dotted line), the current and density polarizations (solid line) calculated at $x=0$  as well as magnetoresistance $R_m$ (dashed-dotted line) 
vs  FM polarizations $p_f$ for $E=1000$V/cm.
}
\label{fig7}
\end{figure}

The lower panel of Fig.~\ref{fig8} illustrates the influence of increasing the FM polarization on the
electrochemical potentials at the junction and in the high field regime ($E=1000$V/cm). 
 Fig.~\ref{fig7} shows in detail the behavior of the related quantities as $p_f$ is increased from 0.5 to full polarization.
 While most of the considered quantities (from the injected current and density polarizations to the sub-band splitting to the magnetoresistance) monotonically increase with the FM polarization to saturate when $p_f=1$, the $\up$
sub-band split (dashed line labelled with $\up$)
increases at first to rapidly drop to zero when the FM polarization saturates. 

As in non-degenerate systems\cite{yu:2002}, $P_0$ saturates towards the $\alpha_0$ value as the field increases (not shown).

\subsection{Magnetoresistance}
A very important quantity for building ``on-off'' type devices as magnetic field sensors or read heads for magnetic hard disk drives\cite{Prinz} is the so called 
magnetoresistance $R_m$. 
It can be defined as the part of the equivalent resistance $R_{eq}$
of the device
 which depends on its magnetic properties.  

For the considered structure $R_{eq}$ is given by
 \begin{eqnarray}\label{reseq1dim}
R_{eq}&\equiv&-{\Delta\mu(\lambda_{sc})-\Delta\mu(-\lambda_{f})\over e|j|},
\end{eqnarray}
 where $\lambda_{sc(f)}$ are the device lengths in the NMS and FM regions,
and $\Delta\mu=(\Delta\mu_\uparrow+\Delta\mu_\downarrow)/2$ indicates 
the variation of the electrochemical potential with respect to the situation in which no electric field is applied, and no junction is 
present.
We will define $R_m$ as  $R_{eq}$ minus the trivial, non-magnetic device resistance $R_{tr}$
when 
$\lambda_{sc}\gg L_{s}$ and $\lambda_{f}\gg L_{f}$\cite{Rtrnote}. 
For a FM/NMS junction we obtain
 \begin{eqnarray}
R_{m}&=& {L_s\over\sigma_{sc}}{L_s\over\mu_s|E|\tau_s}(1-{\rho_{\uparrow\downarrow}\over\rho_D})p_{f}P(0).\label{RMFMNMS}
\end{eqnarray} 
Eq.~(\ref{RMFMNMS}) is very interesting: it clearly shows that $R_m$ depends on the intrinsic resistance of the NMS  $L_s/\sigma_{sc}$, on the ratio between diffusion and drift lengths $L_s/\mu_s|E|\tau_s$ and
it emphasizes the
 dependence of $R_m$ on the spin drag effect.
It also shows that the magnetoresistance is proportional to
the probability $p_{f}P(0)$ that a spin crosses the interface without loosing its spin polarization, i.e. it preserves its magnetic properties when entering the NMS.
Eq.~(\ref{RMFMNMS}) shows that, whichever density regime is considered, the electric field destroys the magnetoresistance.

In Fig.~\ref{fig14} ( lower panel inset, curves labelled ``1jct'') we plot the ratio between $R_m$ 
 and the total device resistance $R_{eq}$ in respect to the applied field and for $\lambda_{f(sc)}=10L_{f(sc)}$: such a ratio increases by orders of magnitude with the FM polarization $p_f$. At  $E$=0.01 V/cm, $R_m=0.28$n$\Omega$cm$^2$ for $p_f=0.5$ and $R_m=8\mu\Omega$cm$^2$ for $p_f=1$. 

\section{FM/NMS/FM structure}
We will now consider the tri-layer structure of Fig.~\ref{fig6.1}b.
According to Eq.~(\ref{deltam}) and to the local charge neutrality constraint
 Eq.~(\ref{chneutr}),
in the NMS the excess spin density components are given by
\begin{equation} \Delta n_{\up(\down)}=\pm\left[A_0\exp(-{x\over L_d})+A_1\exp({(x-x_0)\over L_u})\right]
\end{equation}
where $x_0>0$ corresponds to the second interface and the + (-) sign to the
$\uparrow$ ($\downarrow$) component.
The chemical potentials in the three materials are now 
\begin{eqnarray}
\left(\begin{array}{c}\Delta \mu_\uparrow \\
\Delta \mu_\downarrow
\end{array}\right)_L&=&e{\vec{j}\cdot\vec{x}\over\sigma^f_L} \left(\begin{array}{c}1\\1\end{array}\right)\nonumber \\&+&C_Le\vec{j}\cdot\hat{x}\left(\begin{array}{c}1/\sigma^f_{\uparrow,L} \\-1/\sigma^f_{\downarrow,L}\end{array}\right)\exp({x\over L^f_L})\label{muFM1} \\ \label{muNMS2}
\Delta \mu_{\up(\down)} & =& \pm {D_se\over\mu_s}(1-\rho_{\uparrow\downarrow}/\rho_D)P(x)
+e\vec{E}\cdot \vec{x}+B_{sc}~~\\
\left(\begin{array}{c}\Delta \mu_\uparrow \\\Delta \mu_\downarrow
\end{array}\right)_R&=&e\vec{j}\cdot\hat{x}\left({x\over\sigma^f_R}+B_{f}\right) \left(\begin{array}{c}1\\1\end{array}\right)\nonumber \\&+&C_Re\vec{j}\cdot\hat{x}\left(\begin{array}{c}1/\sigma^f_{\uparrow,R}\\-1/\sigma^f_{\downarrow,R}\end{array}\right)\exp(-{(x-x_0)\over L^f_R})\label{muFM2}
\end{eqnarray}
where the indices $L,~R$ stand for left and right ferromagnets and the + (-) sign in Eq.~(\ref{muNMS2}) corresponds to the $\up$ ($\down$) component.
The six constants $\{A_0,A_1,B_{sc},B_f,C_L,C_R\}$
will be determined by 
boundary conditions similar to the ones for the single interface, i.e.
\begin{eqnarray}
\left.j_s\right|_{0^-,x_0^-}&=&\left.j_s\right|_{0^+,x_0^+} \\
\left.\Delta \mu_\uparrow\right|_{0^-,x_0^-}&=&\left.\Delta \mu_\uparrow\right|_{0^+,x_0^+} \\
\left.\Delta \mu_\downarrow\right|_{0^-,x_0^-}&=&\left.\Delta \mu_\downarrow\right|_{0^+,x_0^+}.
\end{eqnarray}

\begin{figure}
\putfig{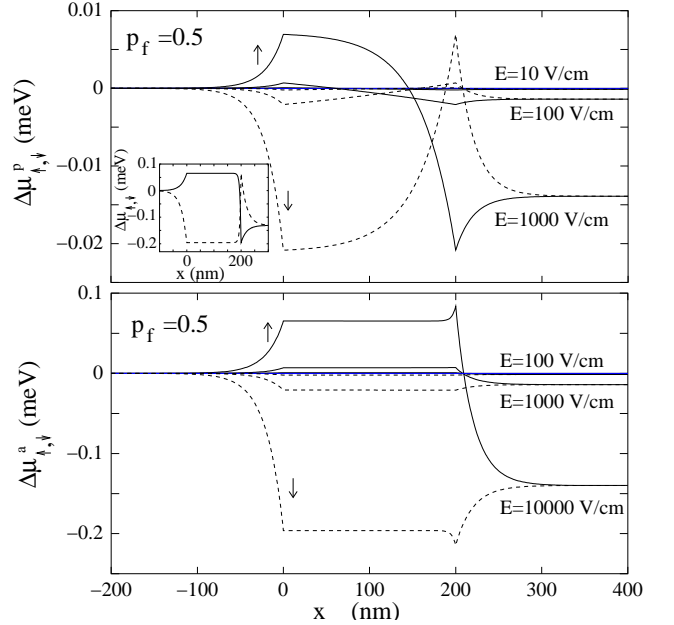}{8.5}
\caption{Upper panel: Up (solid line) and down (dashed line) electrochemical potentials (without the term linear in $x$) vs $x$ for three different fields (as labelled) and $p_f=0.5$. The two ferromagnet polarizations are {\it parallel}.
 Inset: as in main panel but for E=10000 V/cm.
Lower panel:  As in the upper panel but with {\it antiparallel}
ferromagnet polarizations.
}
\label{fig11}
\end{figure}
First of all let us consider
the spin density and current polarizations. 
In order to write the expressions for these quantities we need to derive the expressions for $A_0,~A_1$.
For general polarizations
$p^f_L,~p^f_R$, after some  algebra, the expressions for $A_0,~A_1$  can be written as
\begin{eqnarray}
A_0&=&{n\over2}L_u\left({1\over L_u}-{1\over L_d}\right)\left(p_f^L-\exp(-{x_0\over L_u}){d^-_L\over d^+_R}p_f^R\right)\nonumber\\ & &\cdot \left[u^+_L-\exp(-x_0
\left({1\over L_u}+{1\over L_d}\right))u^-_R{d^-_L\over d^+_R}\right]^{-1}\\
A_1&=&-{n\over2}L_d\left({1\over L_u}-{1\over L_d}\right)\left(p_f^R-\exp(-{x_0\over L_d}){u^-_R\over u^+_L}p_f^L\right)\nonumber\\ & &\cdot \left[d^+_R-\exp(-x_0
\left({1\over L_u}+{1\over L_d}\right))d^-_L{u^-_R\over u^+_L}\right]^{-1},
\end{eqnarray}
where
\begin{eqnarray}d^\pm_{R,L}&\equiv& 1\pm G_{R,L}{L_d\over L^f_{R,L}}\\
u^\pm_{R,L}&\equiv& 1\pm G_{R,L}{L_u\over L^f_{R,L}},
\end{eqnarray} and 
\begin{equation}G_{R,L}\equiv {\sigma^f_{R,L}\over\sigma^{sc}}
[1-(p^{f}_{R,L})^2]
(1-{\rho_{\uparrow\downarrow}\over\rho_D}).
\end{equation}
The spin density and current polarizations are then expressed as
\begin{eqnarray}
P(x)&=&{2\over n}\left[A_0\exp(-{x\over L_d})+A_1\exp({(x-x_0)\over L_u})\right] \\
\alpha(x)&=&{2\over n}\left({1\over L_u}-{1\over L_d}\right)^{-1}\nonumber\\ & &\cdot
\left[A_0\exp(-{x\over L_d}){1\over L_u}-A_1\exp({(x-x_0)\over L_u}){1\over L_d}\right].
\end{eqnarray}
\begin{figure}
\putfig{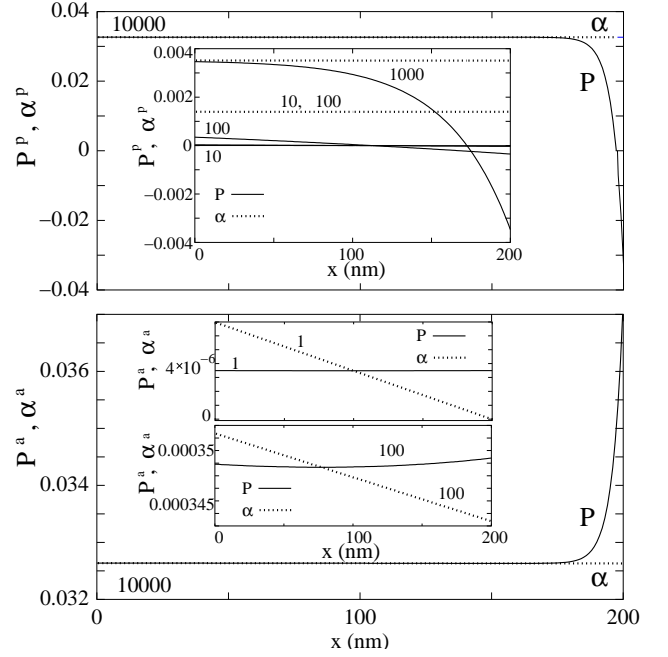}{8.5}
\caption{Upper panel:  density (solid line) and current (dot line) polarizations vs $x$ for E=10000 V/cm and $p_f=0.5$. The two ferromagnet polarizations are {\it parallel}.
Inset: as in main panel but for three different fields (as labelled in units of V/cm). 
Lower panel:  density (solid line) and current (dot line) polarizations vs $x$ for E=10000 V/cm and $p_f=0.5$. The two ferromagnet polarizations are {\it antiparallel}.
Upper inset: as in main panel but for E=1 V/cm.
Lower inset: as in main panel but for E=100 V/cm.
}
\label{fig12}
\end{figure}
In the interesting case in which $p^f_L=\pm p^f_R=p^f$ (parallel and anti-parallel polarization of the two FM's, which are considered, for the rest, equivalent)
 the current polarization
at the two interfaces 
becomes:
\begin{eqnarray}
\alpha(0)&=&{p^f
\over d_R^+u_L^+-\exp(-x_0\left({1\over L_u}+{1\over L_d}\right))d_L^-u_R^-}
\nonumber\\ & &\cdot
\left[d^+_R-\exp(-x_0\left({1\over L_u}+{1\over L_d}\right))u^-_R\right.\nonumber\\ & &\left.\pm\exp(-{x_0\over L_u}){G_L\over L^f_L}(L_u+L_d)\right]\\
\alpha(x_0)&=&{\pm p^f
\over d_R^+u_L^+-\exp(-x_0\left({1\over L_u}+{1\over L_d}\right))d_L^-u_R^-}
\nonumber\\ & &\cdot 
\left[u^+_L-\exp(-x_0\left({1\over L_u}+{1\over L_d}\right))d^-_L\right.\nonumber\\ & &\left.\pm\exp(-{x_0\over L_d}){G_R\over L^f_R}(L_u+L_d)\right],
\end{eqnarray}
where the upper (lower) sign refers to the parallel (antiparallel) configuration.

Finally the expressions for the other constants which appear in the electrochemical potentials are given by
\bea
B_{sc}&=& -SnP(0)p_{L}^f\\
B_f &=& -{x_0\over \sigma^f_R}+{|E|\over|j|}x_0+{Sn\over e|j|}[p^f_LP(0)-p^f_RP(x_0)]\\
C_L&=&-{Sn\over 2}{P(0)\over e|j|}\sigma_L^f(1-p^{f2}_L)\\
C_R&=&-{Sn\over 2}{P(x_0)\over e|j|}\sigma_R^f(1-p^{f2}_R).
\eea 

In Fig.~\ref{fig11} we plot the behavior of the $\up$ (solid lines)
 and $\down$ (dashed lines)  electrochemical potentials in the tri-layer
 structure for the interesting case in which the NMS width $x_0$ is much smaller than the diffusion length ($x_0=L_s/10$). The upper panel shows the electrochemical potentials
when a parallel configuration ($p_{fL}=p_{fR}=0.5$) is considered. 
The lower panel shows the two junction system when the polarizations of the two FM's are opposite (anti-parallel case,$p_{fL}=-p_{fR}=0.5$ ).
For both configurations, as the field is increased the potential drop due to the presence of the NMS increases. 

We notice immediately that in the parallel case the two spin components of the electrochemical potentials cross.  This crossing was predicted to occur in Ref.~\onlinecite{schmidt:2000}
 and is necessary 
 in order to obtain a different potential drop for the $\up$ and $\down$ spin components and hence  a finite current polarization even at low electric fields.
A clear effect of the electric
field enhancement is to ``push'' the crossing between $\up$ and  $\down$ components toward the second interface by increasing the penetration length $L_d$ relative to the first interface and decreasing  the upstream penetration length $L_u$ relative to the second interface.
This behavior becomes extreme for very high fields (see inset where $E=10000$V/cm), where the  electrochemical potential components are almost parallel until
they  suddenly switch before the second interface.

 In the antiparallel configuration (lower panel), while the potential drop across the whole structure behaves similarly to the parallel case, the electrochemical potentials do not cross and this behavior destroys the spin current at very low fields\cite{schmidt:2000}.
 The electrochemical potentials
 present instead a quite distinct feature: for increasing fields 
 a spin accumulation builds up for $x\approx x_0$, 
just before the second interface,
underlining the increasing resistance that spins encounter in crossing this
 interface when the field (and with it the polarization inside the NMS) increases. This anticipates that in the antiparallel configuration the magnetoresistance is expected to increase with increasing applied electric field.

Fig.~\ref{fig12} presents the behavior of the density (solid lines)
and current (dotted line) polarizations as a function of $x$ for both the parallel (upper panel) and the anti-parallel (lower panel) configuration, when the field is increased from the low field regime ($E=1$V/cm, $E=10$V/cm), to the intermediate ($E=100$V/cm), to the high field one ($E=1000$V/cm, $E=10000$V/cm).
The main panel presents the highest field situation.
In both configurations, 
as expected from the behavior in the single junction system,
 for low to intermediate fields $P(0)<\alpha(0)$, while $P(0)\to\alpha(0)$ in the high field regime.
On the other side, in the parallel configuration, 
while, due to the choice $x_0\ll L_s$, $\alpha(x)$ is basically constant even at very low fields,  the density polarization $P(x)$ switches sign
when the two electrochemical potentials cross. This
 inversion in the polarization of the spin population unbalance 
is pushed toward the second interface as the field increases.  

The situation is different in the antiparallel case in which, at low to intermediate fields, an almost constant $P(x)$ corresponds to a constant drop of the 
current polarization $\alpha(x)$ as a function of $x$. For very low fields ($E=1$V/cm), and 
just before the II junction, it becomes even slightly negative, as could be expected by simply considering the diffusion current at this interface.
  The drop in $\alpha(x)$ is due to  the difficulty in establishing a spin current through the system due to the 
abrupt decrease of the $\up$-spin density of states  
in the second ferromagnet.
When the field is increased, the decrement of the current density becomes less drastic while a spin accumulation starts to build up at the second interface (see lower panel of the inset).
 In the high field regime finally, it is possible to establish a basically constant spin density current through the device, at the price of building up a strong spin density accumulation just before the second interface (main lower panel).

\subsection{Magnetoresistance}
Let us now concentrate on the magnetoresistance.
$R_{eq}$ in the FM/NMS/FM structure is given by
 \begin{eqnarray}\label{reseq1}
R_{eq}&\equiv&-{\Delta\mu(x_0+\lambda_{f,R})-\Delta\mu(-\lambda_{f,L})\over e|j|},
\end{eqnarray}
 where the NMS extends from 0 to $x_0$.
By writing $\lambda_{f,R(L)}=nL_{fR(L)}$ and using for the electrochemical potentials Eqs.~(\ref{muFM1}),(\ref{muNMS2}) and (\ref{muFM2}), we obtain
 \begin{eqnarray}\label{reseq}
R_{eq}&=&n\left[{L_{fR}\over\sigma_{fR}}+{L_{fL}\over\sigma_{fL}}\right]+
\left[{x_0\over\sigma_{sc}}\right]\nonumber\\ & & +{L_s\over\sigma_{sc}}(1-\exp(-n)){L_s\over\mu_s|E|\tau_s}(1-{\rho_{\uparrow\downarrow}\over\rho_D})\nonumber \\
& & \cdot [p_{fL}P(0)-p_{fR}P(x_0)]
\end{eqnarray} 
The first line is clearly the trivial, non-magnetic term.
For $n\gg 1$, $R_m$ is given by
 \begin{eqnarray}\label{rm2}
R_{m}&=& {L_s\over\sigma_{sc}}{L_s\over\mu_s|E|\tau_s}(1-{\rho_{\uparrow\downarrow}\over\rho_D})\nonumber \\
& & \cdot [p_{fL}P(0)-p_{fR}P(x_0)]
\end{eqnarray} 
The above equation shows that $R_m$
(i) is proportional to the intrinsic resistance of the paramagnet ${L_s/\sigma_{sc}}$;
(ii) it depends directly on the ratio between the diffusion length $L_s=\sqrt{D_s\tau_s}$ and the drift length $\mu_s|E|\tau_s$,i.e. it tends to diminish with increasing field since the spins will be drifted longer 
and less resistance occurs
(iii) it depends directly on the spin drag which enhances it; 
(iv) it is proportional to the difference $p_{fL}P(0)-p_{fR}P(x_0)=|p_{fL}P(0)|+|p_{fR}P(x_0)|$. This term 
can be seen as the probability that magnetic properties remain relevant through the whole structure.
 It implies that, as long as the products $|p_{f}P|$ are not vanishing,
magnetoresistance is accumulated at the different junctions . 
(v)In the important case in which  $p^f_L=\pm p^f_R=p^f$, the last factor becomes $p_{f}[P(0)\mp P(x_0)]$: this shows that in the anti-parallel case the magnetoresistance is proportional to the sum of the density polarizations just after the first interface and before the second one. This implies that it is 
enhanced by the spin accumulation at the second interface  due to the increasing of the electric field (see Fig.~\ref{fig12}).
\begin{figure}
\putfig{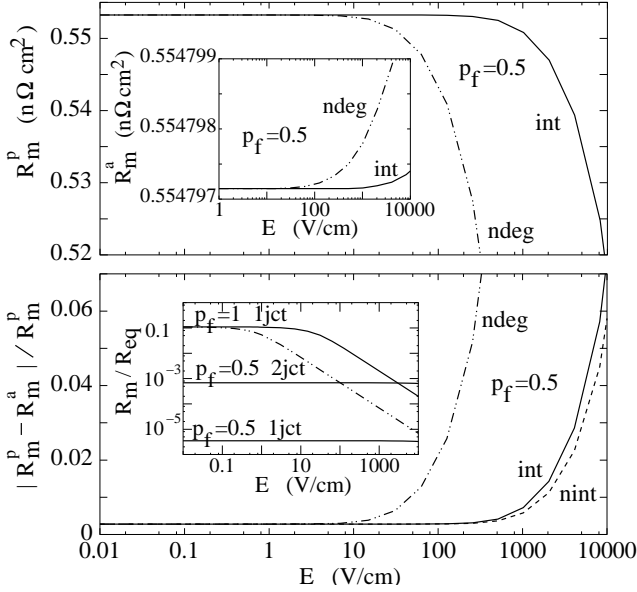}{8.5}
\caption{Upper panel: magnetoresistance $R_m$ for the {\it parallel} configuration
vs electric field for $p_f=0.5$.
Its non-degenerate approximation (dashed-double-dot line) is plotted for comparison. Inset: as for main panel, but for the {\it antiparallel} configuration.
Lower panel: magnetoresistance variation from the the parallel to the antiparallel configuration vs electric field for $p_f=0.5$. Its non-degenerate  (dashed-double-dot line) and non-interacting (dashed line) approximations  are plotted for comparison. Inset: ratio $R_m/R_eq$ vs $E$ for single (1jct) and double junction (2jct, parallel configuration) structures and different FM 
polarizations (as labelled). For $p_f=1$ the dashed-double dot line indicates the non-degenerate approximation.
}
\label{fig14}
\end{figure}

Fig.~\ref{fig14} illustrates the behavior of the magnetoresistance $R_m$ as a function of the electric field. The main upper panel refers to the parallel configuration. We clearly see that the field decreases the magnetoresistance. 
The inset shows the corresponding situation for the {\it antiparallel} case: as anticipated, in this case {\it the magnetoresistance is increased by the applied electric field}. 
The lower panel  presents the ratio of the difference between the parallel $R_m^p$ and the antiparallel $R_m^a$  magnetoresistances to $R_m^p$.
This relative change in magnetoresistance, responsible for the functioning of spin-valve based devices\cite{Prinz},
is increased by the field effect. The non-degenerate and non-interacting approximations are shown for comparison. 
The inset presents the ratio $R_m^p/R_{eq}^p$ (curve labelled ``2jct'') as a function of the applied field. 
It is interesting to notice that this ratio is about two orders of magnitude
 higher than the single junction result with the same FM polarization.
In systems such 
that $L_{sc}/\sigma_{sc}\gg L_{f}/\sigma_{f}$ in fact,
 while the magnetoresistance $R_m$ per junction
 does not change significantly between the 
single and double junction systems, the non-magnetic component of $R_{eq}$
is mainly determined by the NMS layer length, which strongly decreases in the 
the tri-layer device. 
This suggests that, an important requirement for optimizing 
magnetoresistance-based devices, would be to minimize the NMS layer length.
\section{Coulomb interaction effects}
Coulomb interactions enter the drift-diffusion equation
(Eq.~(\ref{steady}))  through the diffusion and mobility constants; in the non-magnetic limit (Eqs.~(\ref{muspin}) and (\ref{Dspin})), through the diffusion constant $D_s$ only, by
affecting the spin-stiffness $S$\cite{PRB:2002} and generating the spin transresistivity $\rho_{\uparrow,\downarrow}$\cite{EPL:2001,PRB:2002,SCD:2000}. 

First of all, we want to point out that 
the experimentally important 
 regime $k_BT\stackrel{<}{\sim}\varepsilon_F$  (see Fig.~\ref{fig1}), is the 
same  in which Coulomb interactions between carriers of opposite 
spin become relevant. 
In fact for $k_BT\sim\varepsilon_F$
the spin transresistivity reaches its maximum and can become of the same order of the Drude resistivity\cite{PRB:2002}, while for $k_BT\stackrel{<}{\sim}\varepsilon_F$  the spin-stiffness displays the maximum deviation from its
 non-interacting approximation\cite{PRB:2002}. These two combined effects 
{\it reduce} the  non-interacting approximation to $D_s$ even by 50\%, depending on carrier density and temperature\cite{PRB:2002}. According to  the functional  dependence on $D_s$ of the various quantities of interest,
Coulomb interaction  effects can  get amplified, 
as we will explicitly show 
for the downstream 
spin penetration length and for the spin current.

\begin{figure}
\putfig{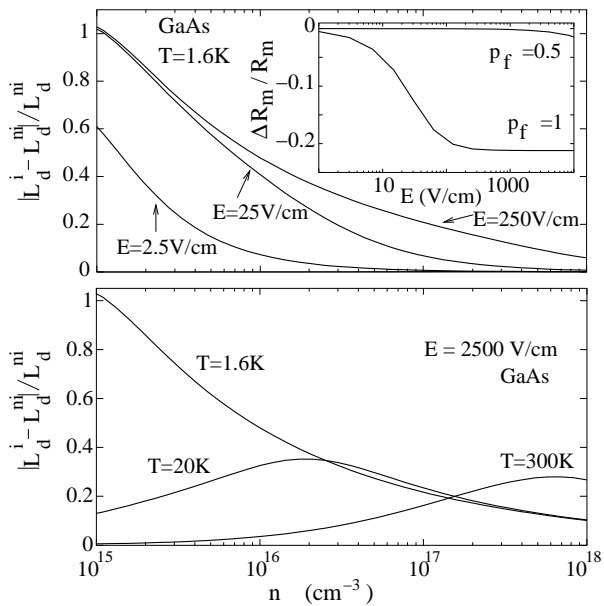}{8.0}
\caption{
Upper panel:   $|L_d^i-L_d^{ni}|/L_d^{ni}$ vs carrier density for GaAs parameters, T=1.6K and three different electric fields (as labelled). Inset: 
 $\Delta R_m/R_m\equiv
(R_m^i-R_m^{ni})/R_m^{ni}$  vs electric field for $p_f=0.5$ and $p_f=1$ (as labelled). 
Lower panel:   $|L_d^i-L_d^{ni}|/L_d^{ni}$ vs carrier density for GaAs parameters, T=1.6K , 20K  and 300K (as labelled) and E=2500V/cm. 
}
\label{fig15a}
\end{figure}

The strong contribution of Coulomb interactions to $L_{u,d}$ values
 is underlined in
the upper panel of Fig.~\ref{fig15a} (as well as by the direct comparison between solid and dashed lines in  Fig.~\ref{fig2})
 in which
 we plot, for GaAs at low temperature, 
 the correction  $(L_{d}^{i}-L_d^{ni})/L_d^{ni}$ as a function of carrier density and for different fields.
Here and in the following $i,ni$ stand for interacting and non-interacting respectively. 
As the figure shows, the correction can be of the order of 100\%, definitely not-negligible when quantitative calculations are required. 
This correction becomes the largest in the ``semiconductor degenerate regime''\cite{note4} $e|E|L_s>>\varepsilon_F>>k_BT$, when the penetration lengths approach their high field limits  $L_d\approx L_s^2E\mu_s/D_s$, $L_u\approx D_s/(E\mu_s)$, which
strongly depend over the diffusion constant.
At low densities such a limiting behavior is reached already for $E=25$V/cm, while the curve corresponding to $E=2.5$V/cm shows smaller corrections.
For higher densities the limiting behavior is reached at larger fields
 (see the curve corresponding to $E=250$V/cm), though the interacting correction remains significative even for fields as low as $E=25$V/cm.

The lower panel  of Fig.~\ref{fig15a} shows 
 the importance of Coulomb corrections in the high field regime, when temperature and carrier density are varied. 
The largest corrections are present at   $T=1.6$K and low densities and are due to the spin stiffness reduction.
 Even at temperatures as high as room temperature though, Coulomb interactions remain relevant and the correction can be 
as high as 30\%. We underline that in this last case, it is  mainly due to
 the spin Coulomb drag (SCD) effect\cite{SCD:2000,PRB:2002} which enters the diffusion constant through the spin transresistivity (see Eq.~(\ref{Dspin})).
We want to stress that Coulomb interactions modify the behavior of $L_{u,d}$ even qualitatively: as shown for $L_d$ in the lower panel of Fig.~\ref{fig2} 
 ($T=20,~300$K), due to Coulomb corrections, the non-degenerate limit is approached from above instead than from below when $n$ is decreased. This implies that, even if for high enough temperatures the system can actually enter the non-degenerate regime,
 the maximum of $L_d$ (the minimum of  $L_u$) is in any case  
reached {\it outside} such a  regime.

The inset of Fig.~\ref{fig5} shows the Coulomb correction to $j_D^d$ as a function of carrier density and for three different electric fields (as labelled in units of V/cm). Again we notice
 that the functional dependence of $j_D^d$ on $D_s$
 amplifies interaction effects which,
depending on the carrier density,
 are quite substantial. 

Let us now consider how Coulomb interactions should affect  heterostructures
 between materials with different magnetic properties. Apart from entering 
the equations of motion through quantities as the diffusion constant or the
penetration lengths, we expect the SCD to play a special role
and to affect independently all the quantities related to 
interface properties. 
This is due to the very nature of the SCD effect:  in fact it tends to relax spin current {\it opposing relative motion} between up and down spin components,
 which include opposing all their relative variations connected to  injection of
 a drifting spin unbalance.  
This picture is confirmed by all our equations related to
 heterostructures, starting from the expression for the electrochemical potential Eq.~(\ref{deltamuNMS}), in which
the spin Coulomb drag enters
 explicitly through the factor $(1-\rho_{\uparrow\downarrow}/\rho_D)$. 
As a consequence, all quantities derived from this potential as injected spin density, current density, and magnetoresistance,
will  explicitly display such a factor.

The spin-transresistivity $\rho_{\uparrow\downarrow}$ is a negative quantity,
so the factor $(1-\rho_{\uparrow\downarrow}/\rho_D)$   will {\it oppose spin injection} in
 Eqs.~(\ref{P0}) and  (\ref{alpha0}), while it will {\it increase the magnetoresistance} in Eqs.~(\ref{Rm1jct0}) and
(\ref{rm2}). Since for some systems $\rho_{\uparrow\downarrow}\approx \rho_D$\cite{PRB:2002}, the SCD could  actually strongly affect the magnetoresistance.
In the calculations here presented though, we have considered only degenerate systems, in which the SCD  is usually negligible.

In the inset of Fig.~\ref{fig8}, for   $E=1000$V/cm, we plot both the interacting (solid line) and the
non-interacting  (dashed line) calculations of 
the electrochemical potentials: as 
early discussed, at high fields, interactions increase
 the spin penetration length $L_d$ in a noticeable
 way; this  correspondingly increases
 the extension of the polarized region in the NMS.

\begin{figure}
\putfig{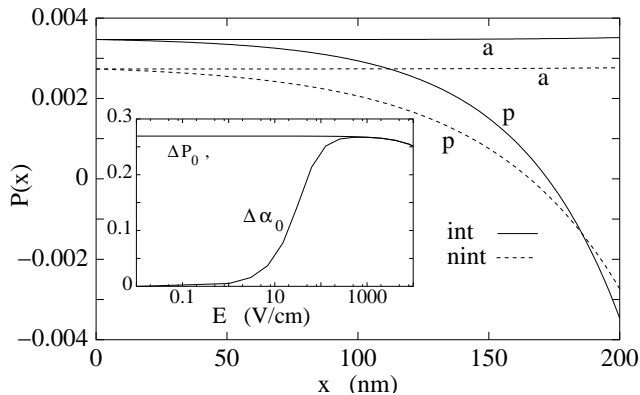}{8.5}
\caption{ Double junction system: Density polarization $P(x)$ vs $x$ for the parallel (labelled as ``p'') and antiparallel (``a'') configurations (solid line). The non-interacting approximation (dashed line) is plotted as well. Inset: Single junction system:
  $\Delta P_0\equiv (P_0^i-P_0^{ni})/P_0^{ni}$ and $\Delta \alpha_0\equiv (\alpha_0^i-\alpha_0^{ni})/\alpha_0^{ni}$  vs electric field.
}
\label{fig13}
\end{figure}

For all calculations related to  heterostructures,
 we have considered the carrier density $n=5.2\times 10^{16}$cm$^{-3}$:
as can be expected by the upper panel of Fig.~\ref{fig15a}, the 
influence of Coulomb interactions should then be not negligible and of the order of 20-25\%.
This is confirmed by the detailed analysis of the various quantities of interest:
in a FM/NMS heterostructure
 the 
influence of Coulomb interactions on $\alpha(0)$ and $P(0)$   is  of the order of 25\% (see inset of Fig.~\ref{fig13}). 
For $E\to 0$, the ratio $(P_0^i-P_0^{ni})/P_0^{ni}\to(D_s^i-D_s^{ni})/D_s^{ni}$, while 
as the field is increased well inside the high field regime,
$L_u$ strongly decreases and the Coulomb related term in Eqs.~(\ref{P0}) and
(\ref{alpha0}) becomes less important, so that the Coulomb correction starts to decrease.
For the same structure, the inset  of Fig.~\ref{fig15a} presents  the Coulomb interaction effects on $R_m$  for increasing electric field and polarization. Here $\Delta R_m/R_m\equiv
(R_m^i-R_m^{ni})/R_m^{ni}$. 
We see that  such effects become important as both field and $p_f$ increases.

When we consider a double-junction heterostructure, the effect of Coulomb 
interactions remains sizeable. 
In Fig.~\ref{fig13} we plot  the interacting (solid line) and non-interacting (dashed line) spin  density polarization $P(x)$ across the NMS, both
 in the parallel (labelled by ``p'') and antiparallel (labelled by ``a'') configuration for $E=1000$V/cm.
 Again, as for the one-junction case, the interaction correction
is of the order of 20-25\%.
Similarly from the lower panel of Fig.~\ref{fig14} we see that Coulomb interactions at high fields and at the considered density 
can influence the ratio $(R_m^p-R_m^a)/R_m^p$ 
by the same amount.

\section{Conclusions}
We have presented a detailed study of spin injection into
 degenerate semiconductors,  analyzing the combined effects of applied electric field, carrier density and Coulomb interactions.

We have clarified the interplay of the different energy scales involved in the problem, proposing an efficient way to understand the regime of the system and estimate the effect of the applied electric field.
In particular we have focused on ferromagnet/non-magnetic semiconductor and ferromagnet/non-magnetic semiconductor/ferromagnet structures, carefully comparing the behavior of the relevant quantities (electrochemical potentials, spin density, magnetoresistance...) when the polarization of the second ferromagnet is switched from parallel to antiparallel or its absolute value and the electric field are increased. 

We have pointed out, in the anti-parallel case, that the presence of a spin-current through the device is related to the onset of a spin accumulation at the second interface, which  in turn increases the device magnetoresistance: at variance with the parallel case, we demonstrate that in the anti-parallel configuration the magnetoresistance can be increased by a 
 strong applied electric field. Additionally we show that this field increases as well the relative change in magnetoresistance $|R_m^p-R_m^a|/R_m^p$, exploited in spin-valve based metal devices\cite{Prinz}. 
  
We have proposed a general formalism which includes Coulomb interactions
and applies to FM/NMS junctions
 independently of their carrier density regime.
  In particular it allows also to study  heterostructures in the intermediate regime $\varepsilon\approx k_BT$.
``Exact'' expressions for the case in which a non-interacting degenerate system is considered have also been  provided (see Appendix B).

Our calculations show that, in the experimentally important regime $\varepsilon_f\stackrel{>}{\sim}k_BT$, corrections due to Coulomb interactions
are relevant. They also demonstrate the importance of carefully identifying the system regime 
(for example by comparing energy scales as we suggest in Sec.II): non-degenerate approximations in fact, when extended to  the wrong regime,
can lead to errors of orders of magnitude.
\section {Acknowledgements}We thank L. W. Molenkamp and G. Schmidt for helpful discussions, G. Vignale for for several discussions and very constructive comments and P.E. Mason for carefully reading the manuscript. 
\appendix
\section{Limit of the chemical potential approximation}
The approximation  we proposed for the chemical potential, Eq.~(\ref{muappr}),
can be thought at first sight to be valid only in the low polarization limit $P(x)\ll 1$.

In this appendix we want to estimate the limits of this approximation 
in the important case of a degenerate system.

\begin{figure}
\putfig{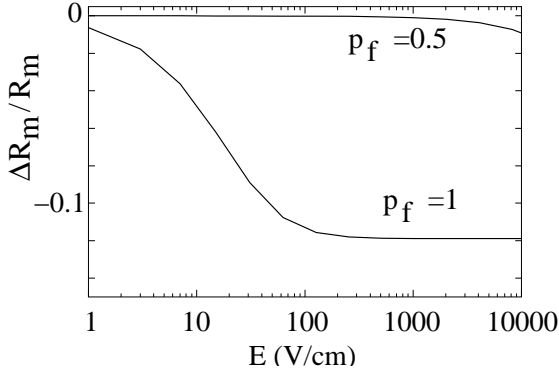}{7.5}
\caption{Correction to the (non-interacting)
degenerate approximation for $R_m$, $\Delta R_m/R_m\equiv
(R_m^{exa}-R_m^{deg})/R_m^{deg}$  vs electric field for $p_f=0.5$ and $p_f=1$ (as labelled).
Here $R_m^{exa}$ corresponds to the
(non-interacting)
 ``exact'' degenerate calculation.
}
\label{fig16}
\end{figure}

In a non-interacting system the Taylor expansion Eq.~(\ref{Taylor}) becomes
  \begin{equation}
\mu_\sigma^{chem}=\mu_{0\sigma}^{chem}+
{\partial\mu_\sigma^{chem}\over\partial n_\sigma}\Delta n_\sigma
+{1\over 2}{\partial^2\mu_\sigma^{chem}\over\partial n_{\sigma}^2}\Delta n_{\sigma}^2+.....\label{Taylor2}
\end{equation}
In a degenerate non-magnetic system, the chemical potential (up to terms of order $T^2$) is given by
\be
\mu_\sigma^{chem}=\varepsilon_{F\sigma}={\hbar^2\over 2m}(6\pi^2n_\sigma)^{2/3}\label{mudeg}
\ee
Truncating Eq.~(\ref{Taylor2}) up to first order, is equivalent to request
\be
{1\over 2}\left|{\partial^2\mu_\sigma^{chem}\over\partial n_{\sigma}^2}\Delta n_{\sigma}^2\right|\ll \left|{\partial\mu_\sigma^{chem}\over\partial n_\sigma}\Delta n_\sigma\right|,
\ee
which, using Eq.~(\ref{mudeg}) becomes
\be
P(x)\ll 6 \label{Plimit}.
\ee
The above equation implies
 that,  in the important case of a degenerate system, our approximation is valid up to injected polarizations of at least 
$P(x)\sim 0.5$\cite{note6}.
In the special case of a single junction heterostructure it remains valid up to density polarizations of the order of 1.
This  is shown in Fig.~\ref{fig16},  where the effects of the chemical potential approximation  discussed at the beginning of Sec.IV are presented
for increasing electric field and FM polarization.
Fig.~\ref{fig16}
compares the results obtained for $R_m$ in the FM/NMS structure
using the non-interacting 
(degenerate) approximation to the chemical potential, with  
the corresponding
``exact'' non-interacting degenerate results (see Appendix B).
The figure shows that our approximation for the chemical potential gives reasonably small errors (at most of the order of 10\%) even for $p_f=1$. We caution though that this feature cannot be blindly generalized. 

Since $|P(x)|\le p_f$ and the equality corresponds to the  high field limit,
 we have analyzed systems in which $p_f$ satisfies  
the same limitations found for $P(x)$.

\section{Non-interacting degenerate system}
If we consider a non-interacting, degenerate NMS,
Eq.~(\ref{deltamugen}) for the excess electrochemical potential becomes  
\bea
\Delta \mu_\sigma&=&e\vec{E}\cdot \vec{x}+(\varepsilon_{F\sigma}-\varepsilon_{F\sigma}^0)+B\\ 
&=& e\vec{E}\cdot\vec{x}+\varepsilon_{F\sigma}^0\left[\left(1\pm P(x)\right)^{2/3}-1\right]+B
\eea
where the +(-) sign corresponds to $\sigma=\uparrow~(\downarrow)$ and $\varepsilon_{F\sigma}^0$ is the Fermi energy of the unperturbed system.

Applying the conditions (\ref{cond1})-(\ref{cond3}), we can now 
derive the equation for the injected density polarization P(0)
\bea
&-&{eL_f\over 2 \varepsilon_{F\sigma}^0}(|j|+|j_D|){4\over\sigma_f}{1\over1-p_f^2}P(0)-\{[1+P(0)]^{2/3}\nonumber\\&-& [1-P(0)]^{2/3}\}+{e|j|L_f\over 2 \varepsilon_{F\sigma}^0}p_f{4\over\sigma_f}{1\over1-p_f^2}=0.
\eea
Once $P(0)$ is known it is easy to derive all other quantities of interest.

A similar scheme can be applied to the two junction system, where 
we obtain a system of two implicit equations for the two quantities $P(0)$ and $P(x_0)$, which must be solved self consistently .


\begin{thebibliography}{100}
\bibitem{note4}The interacting and non-interacting systems are defined as having the same diffusion length $L_s$ when no external field is applied.
\bibitem{Rtrnote}In systems, for which $L_{sc}/\sigma_{sc}\gg L_{f}/\sigma_{f}$, the non-magnetic resistance is dominated by the semiconductor component, i.e. $R_{tr}\sim \lambda_{sc}/\sigma_{sc}$.
\bibitem{note6}If we apply the same analysis to the non-degenerate case, where $\mu_\sigma^{chem}\sim \ln n_\sigma$, we obtain $P(x)\ll 2$.
\end{thebibliography}
\end{document}